\documentclass[preprint,floatfix,showpacs]{revtex4}%
\usepackage{graphicx}
\usepackage{amsmath,amssymb}
\usepackage{amsmath}
\usepackage{amsfonts}
\usepackage{amssymb}%
\def\c5q{c^5_q}

\def\pl2{\ell^2}
\def\pq2{q^2}

\begin{document}
\title{Pion radiative weak decays in nonlocal chiral quark models}
\author{D. G\'{o}mez Dumm$^{a,b}$, S. Noguera$^{c}$ and N. N. Scoccola$^{b,d,e}$}
\affiliation{$^{a}$ IFLP, CONICET $-$ Dpto.\ de F\'{\i}sica, Universidad Nacional de La
Plata, C.C. 67, (1900) La Plata, Argentina.}
\affiliation{$^{b}$ CONICET, Rivadavia 1917, (1033) Buenos Aires, Argentina.}
\affiliation{$^{c}$ Departamento de F\'{\i}sica Te\'orica and Instituto de F\'{\i}sica
Corpuscular, Universidad de Valencia-CSIC, E-46100 Burjassot (Valencia), Spain.}
\affiliation{$^{d}$ Physics Department, Comisi\'on Nacional de Energ\'{\i}a At\'omica, }
\affiliation{Av.\ Libertador 8250, (1429) Buenos Aires, Argentina}
\affiliation{$^{e}$ Universidad Favaloro, Sol{\'\i}s 453, (1078) Buenos Aires, Argentina.}

\begin{abstract}
We analyze the radiative pion decay $\pi^{+}\rightarrow e^{+}\nu_{e}\gamma$
within nonlocal chiral quark models that include wave function
renormalization. In this framework we calculate the vector and axial-vector
form factors $F_V$ and $F_A$ at $q^2=0$ ---where $q^2$ is the $e^+\nu_e$
squared invariant mass--- and the slope $a$ of $F_V(q^2)$ at $q^2\to 0$. The
calculations are carried out considering different nonlocal form factors, in
particular those taken from lattice QCD evaluations, showing a reasonable
agreement with the corresponding experimental data. The comparison of our
results with those obtained in the (local) NJL model and the relation of
$F_V$ and $a$ with the form factor in $\pi^0\to\gamma^*\gamma$ decays are
discussed.
\end{abstract}

 \maketitle

The radiative pion decay $\pi^{+}\rightarrow e^{+}\nu_{e}\gamma$
is a very interesting process from different points of view.
According to the standard description, the corresponding decay
amplitude consists of the inner bremsstrahlung (IB) and
structure-dependent (SD) terms. The IB contribution corresponds to
the situation in which the photon is radiated by the electrically
charged external legs, either pion or lepton, while the SD terms
are associated with the photon emission from intermediate states
generated by strong interactions. The latter can be
parameterized through the introduction of vector and axial-vector
form factors, $F_V(q^2)$ and $F_A(q^2)$ respectively, where $q^2$
is the squared invariant mass of the $e^+\nu_e$
pair~\cite{Bryman:1982et}. Since $\pi^{+}\rightarrow
e^{+}\nu_{e}$ is helicity suppressed, same happens to the IB
contribution to its radiative counterpart, and, consequently,
$\pi^{+}\rightarrow e^{+}\nu_{e}\gamma$ turns out to be an
appropriate channel to uncover the nonperturbative SD amplitude.
Recent measurements\cite{Bychkov:2008ws} of the $\pi^+ \rightarrow
e^+ \nu_e \gamma$ branching ratio over a wide region of phase
space yields $F_V(0) = 0.0258(17)$, $F_A(0)=0.0117(17)$ and $a=
0.10(6)$, where $a$ is related to the dependence of $F_V$ on
$q^2$ parameterized as
\begin{equation}
F_{V}(q^{2})\ \simeq \ F_{V}(0)\left(1+a \ q^2/m_{\pi^+}^2\right)
\qquad {\rm for} \ \ q^{2}\ll m_{\pi^+}^{2}\ .
\label{fv}
\end{equation}
From the point of view of the physics of weak interactions, this
determination has provided a further check of
conserved-vector-current (CVC) hypothesis. In fact, the value of
$F_V(0)$ given above is in good agreement to that extracted from
the analysis of the $\pi^0 \rightarrow \gamma \gamma$ decay
\cite{pdg}. Moreover, as stressed in Ref.~\cite{Bychkov:2008ws}, a
good description of the data appears to be possible without the
need to include extra tensor contributions that arise in several
extensions of the Standard Model suggested in the
literature~\cite{chiz,pobla}.  From the side of strong interaction
physics, the pion radiative decays have been analyzed using Chiral
Perturbation Theory~\cite{Hol86} and effective meson lagrangian
methods~\cite{Mateu:2007tr}. However, there still remains the
question of how the associated form factors are connected to the
underlying quark structure. Due to the nonperturbative nature of
the quark-gluon interactions in the low-energy domain, to address
this issue one is forced to deal with models that treat quark
interactions in some effective way. Among these, the
Nambu--Jona-Lasinio (NJL) model has been widely used as an
schematic effective theory for
QCD~\cite{Vogl:1991qt,Klevansky:1992qe,Hatsuda:1994pi}, allowing
e.g.\ the description of light mesons as fermion-antifermion
composite states. In the NJL model quarks interact through a
local, chiral invariant four-fermion coupling. The corresponding
predictions for the vector and axial-vector form factors at
$q^{2}=0$ have been calculated in Ref.~\cite{Courtoy:2007vy},
yielding $F_{V}(0)=0.0244$ and $F_{A}(0)=0.0241$. An extension of
that calculation leads to $a=0.032$. As we see, while the
predictions in the vector sector are in reasonable agreement with
the measured values (the prediction for $a$ being slightly below
the empirical range), the calculated value of $F_{A}(0)$ is a
factor 2 too large. As a way to improve upon the NJL model,
extensions which include nonlocal interactions have been proposed
(see Ref.~\cite{Rip97} and references therein). In fact,
nonlocality arises naturally in several well established
approaches to low energy quark dynamics. This is e.g.\ the case of
the instanton liquid model~\cite{SS98} and the Schwinger-Dyson
resummation techniques~\cite{Roberts:1994dr}, and also lattice QCD
calculations~\cite{Parappilly:2005ei,Bowman2003,Furui:2006ks}
indicate that quark interactions should act over a certain range
in momentum space. Indeed, nonlocal chiral quark models have been
successfully applied to study different hadron observables
\cite{Bowler:1994ir,Scarpettini:2003fj,GomezDumm:2006vz,Golli:1998rf,Rezaeian:2004nf,Praszalowicz:2001wy,Noguera,Noguera:2005cc,Noguera:2008}.
The aim of the present work is to investigate the predictions of
this type of models for the measured quantities associated to
vector and axial-vector form factors in $\pi^{+}\rightarrow
e^{+}\nu_{e}\gamma$ decays.

In general, the amplitude for the process $\pi^{+}\rightarrow
e^{+}\nu_{e} (q)+\gamma(k)$ can be written as~\cite{Bryman:1982et}
\begin{eqnarray}
\mathcal{M} &=& \frac{G_{F}}{\sqrt{2}}\ e\,\cos\theta_{C}\;\varepsilon_{\mu}
\left[
\sqrt{2}\;f_{\pi}\left\{  (q + k)^{\alpha}\,L_{\alpha\mu}%
\;-\;l^{\nu}\left[  g_{\mu\nu}\;+\;\frac{q_{\mu}q_{\nu}}{(q\cdot k)}\right]
\right\}\right.
\nonumber \\
& & \left. \qquad \qquad\qquad
+ l^{\nu}\left\{  -\,i\,\epsilon_{\mu\nu\alpha\beta}\;k^{\alpha
}q^{\beta}\;\frac{F_{V}(q^{2})}{m_{\pi}}\;+\left[  q_{\mu}k_{\nu}%
\;-\;g_{\mu\nu}\,(q\cdot k)\right]  \frac{F_{A}(q^{2})}{m_{\pi}}\right\} \
\right]  \ ,
\label{dos}
\end{eqnarray}
with $(q+k)^{2}=m_{\pi}^{2}$, $k^{2}=0$. Here $G_F$ and $\theta_C$ stand
for  the Fermi constant and the Cabibbo angle, respectively;
$\varepsilon_{\mu}$ is the photon polarization vector, $l_{\mu}$ is the
lepton current, $L_{\alpha\mu}$ is a lepton tensor, and $F_{V}(q^{2})$ and
$F_{A}(q^{2})$ denote the vector and axial-vector hadronic form factors
mentioned above. In this work we are interested in the study of the
predictions for the measured quantities associated with these form factors
in the context of  nonlocal chiral models. We consider SU(2) chiral
models that include wave function renormalization, defined by the
following Euclidean action \cite{Noguera,Noguera:2008}
\begin{equation}
S_{E}=\int d^{4}x\ \left\{  \bar{\psi}(x)\left(  -i\rlap/\partial
+m_{c}\right)  \psi(x)-\frac{G_{S}}{2}\Big[j_{a}(x)j_{a}(x)+j_{P}%
(x)j_{P}(x)\Big]\right\}  \ . \label{action}%
\end{equation}
Here $m_{c}$ is the current quark mass, which is assumed to be equal for $u$
and $d$ quarks, while the nonlocal currents $j_{a}(x),j_{P}(x)$ are given by
\begin{align}
j_{a}(x)  &  =\int d^{4}z\ g(z)\ \bar{\psi}\left(  x+\frac{z}{2}\right)
\ \Gamma_{a}\ \psi\left(  x-\frac{z}{2}\right)  \ , \nonumber\\
j_{P}(x)  &  =\int d^{4}z\ f(z)\ \bar{\psi}\left(  x+\frac{z}{2}\right)
\ \frac{i{\overleftrightarrow{\rlap/\partial}}}{2\ \varkappa_{p}}\ \psi\left(
x-\frac{z}{2}\right)\ ,
\label{cuOGE}
\end{align}
where $\Gamma_{a}=(\leavevmode\hbox{\small1\kern-3.8pt\normalsize1}, i
\gamma_{5}\vec{\tau})$ and $u(x^{\prime}) {\overleftrightarrow{\partial}}
v(x)=u(x^{\prime})\partial_{x}v(x) - \partial_{x^{\prime}}
u(x^{\prime})v(x)$. The functions $g(z)$ and $f(z)$ in Eq.~(\ref{cuOGE})
are nonlocal covariant form factors characterizing the corresponding
interactions. In what follows it is convenient to Fourier transform $g(z)$
and $f(z)$ into momentum space. Note that Lorentz invariance implies that
the Fourier transforms $g_{p}$ and $f_{p}$ can only be functions of
$p^{2}$.

In order to deal with meson degrees of freedom, one can perform a standard
bosonization of the theory. This is done by considering the corresponding
partition function $\mathcal{Z}=\int\mathcal{D}\bar{\psi}\,\mathcal{D}%
\psi\,\exp[-S_{E}]$, and introducing auxiliary fields $\sigma_{1}%
(x),\sigma_{2}(x),\vec{\pi}(x)$, where $\sigma_{1,2}(x)$ and $\vec{\pi}(x)$
are scalar and pseudoscalar mesons, respectively. An effective bosonized
action is obtained once the fermion fields are integrated out. To treat that
bosonic action we assume, as customary, that $\sigma_{1,2}$ fields have
nontrivial translational invariant mean field values $\bar{\sigma}_{1,2}$,
while the mean field values of pseudoscalar fields $\pi_{i}$ are zero.
Thus we write
\begin{equation}
\sigma_{1}(x)=\bar{\sigma}_{1}+\delta\sigma_{1}(x)\ ,\qquad\sigma
_{2}(x)=\varkappa_{p}\ \bar{\sigma}_{2}+\delta\sigma_{2}(x)\ ,\qquad
\vec{\pi}(x)=\delta\vec{\pi}(x) \ .
\end{equation}
Replacing in the bosonized effective action and expanding in powers of
meson fluctuations we get
\[
S_{E}^{\mathrm{bos}}\ =\ S_{E}^{\mathrm{MFA}}\ + \ S_{E}^{\mathrm{quad}}\
+ \ ...
\]
Here the mean field action per unit volume reads
\begin{equation}
\frac{S_{E}^{\mathrm{MFA}}}{V^{(4)}}=\frac{1}{2G_{S}}\left(  \bar{\sigma}%
_{1}^{2}+\varkappa_{p}^{2}\ \bar{\sigma}_{2}^{2}\right)  -4N_{c}\int
\frac{d^{4}p}{(2\pi)^{4}}\ \ln\left[  \frac{z_{p}}{-\rlap/p+m_{p}}\right]
^{-1}\ ,
\end{equation}
with
\begin{equation}
z_{p}=\left(  1-\bar{\sigma}_{2}\ f_{p}\right)^{-1}\ ,\qquad m_{p}
=z_{p}\left(  m_{c}+\bar{\sigma}_{1}\ g_{p}\right)\ .
\label{mz}
\end{equation}
The minimization of $S_{E}^{\mathrm{MFA}}$ with respect to $\bar{\sigma}%
_{1,2}$ leads to the corresponding gap equations. The quadratic terms can be
written as
\begin{equation}
S_{E}^{\mathrm{quad}}=\frac{1}{2}\int\frac{d^{4}p}{(2\pi)^{4}}\sum
_{M=\sigma,\sigma^{\prime},\pi}G_{M}(p^{2})\ \delta M(p)\ \delta M(-p)\ ,
\label{quad}%
\end{equation}
where $\sigma$ and $\sigma^{\prime}$ fields are meson mass eigenstates,
defined in such a way that there is no $\sigma-\sigma^{\prime}$ mixing at
the level of the quadratic action. The explicit expressions of
$G_{M}(p^{2})$, as well as those of the gap equations mentioned above, can
be found in Ref.~\cite{Noguera:2008}. Meson masses can be obtained by
solving the equation $G_{M}(-m_{M}^{2})=0$, while on-shell meson-quark
coupling constants $g_{Mq\bar{q}}$ are given by
\begin{equation}
{g_{Mq\bar{q}}}^{-2}=\ \frac{dG_{M}(p^2)}{dp^{2}}\bigg|_{p^{2}=-m_{M}^{2}}\ .
\label{gpiqq}%
\end{equation}
As in Ref.~\cite{Noguera:2008}, we will consider here different functional
dependencies for the form factors $g_{p}$ and $f_{p}$. First, we consider a
relatively simple case in which there is no wave function renormalization of
the quark propagator, i.e.~$f_{p}=0$, $z_{p}$ = 1, and we take an often used
exponential parameterization for $g_{p}$,
\begin{equation}
g_{p}=\mbox{exp}\left(  -p^{2}/\Lambda_{0}^{2}\right)\ .
\label{fg}
\end{equation}
The model parameters $m_c$, $G_S$ and $\Lambda_0$ are determined by
fitting the pion mass and decay constant to their empirical values
$m_{\pi}=139$ MeV and $f_{\pi }=92.4$ MeV, and fixing the chiral
condensate to the phenomenologically acceptable value
$\langle\bar{q}q\rangle^{1/3} = -240$ MeV. In what follows we refer to this
choice of model parameters as Set A. Second, we consider a more general
case that includes the wave function renormalization of the quark
propagator. We keep the exponential shape (\ref{fg}) for the form factor
$g_p$ and assume also an exponential form for $f_p$, namely
\begin{equation}
f_{p}=\mbox{exp}\left(  -p^{2}/\Lambda_{1}^{2}\right)\ .
\label{ff}
\end{equation}
Note that the range (in momentum space) of the nonlocality in each channel
is determined by the parameters $\Lambda_{0}$ and $\Lambda_{1}$,
respectively. As in the previous case, model parameters are determined so as
to reproduce the desired values of $m_{\pi}$, $f_{\pi}$ and
$\langle\bar{q}q\rangle^{1/3}$. The form factor $f_p$ introduces now an
additional free parameter $\Lambda_1$, consequently we introduce as a fourth
requirement the condition $z_{p}(0)=0.7$, which is within the
range of values suggested by recent
lattice calculations~[8, 10]. This choice of model parameters and form
factors will be referred to as parameterization Set B. Finally, we consider
a different functional form for the form factors, given by
\begin{equation}
g_{p}=\frac{1+\alpha_{z}}{1+\alpha_{z}\ f_{z}(p)}\frac{\alpha_{m}%
\ f_{m}(p)-m\ \alpha_{z}f_{z}(p)}{\alpha_{m}-m\ \alpha_{z}}\ ,\qquad
f_{p}=\frac{1+\alpha_{z}}{1+\alpha_{z}\ f_{z}(p)}f_{z}(p)\ ,
\end{equation}
where
\begin{equation}
f_{m}(p)=\left[  1+\left(  p^{2}/\Lambda_{0}^{2}\right)^{3/2}\right]
^{-1}\ ,\qquad f_{z}(p)=\left[  1+\left(  p^{2}/\Lambda_{1}^{2}\right)
\right]^{-5/2}\ . \label{parametrization_set2}%
\end{equation}
As shown in Ref.~\cite{Noguera:2008}, taking $m_{c}=2.37$ MeV, $\alpha
_{m}=309$ MeV, $\alpha_{z}=-0.3$, $\Lambda_{0}=850$ MeV and $\Lambda_{1}%
=1400$~MeV one can very well reproduce the momentum dependence of mass and
renormalization functions obtained in lattice calculations, as well as the
physical values of $m_{\pi}$ and $f_{\pi}$. In what follows we will refer to
this choice of model parameters as parameterization Set~C. The parameter
values for all three parameter sets, as well as the corresponding
predictions for several meson properties, can be found in
Ref.~\cite{Noguera:2008}.

In order to derive the form factors we are interested in, one should
\textquotedblleft gauge\textquotedblright\ the effective action $S_{E}$ by
introducing the electromagnetic field $A_{\mu}(x)$ and the charged weak
fields $W_{\mu }^{\pm}(x)$. For a local theory this \textquotedblleft
gauging\textquotedblright\ procedure is usually done by performing the
replacement
\begin{equation}
\partial_{\mu}\rightarrow\partial_{\mu}+i\ G_{\mu}(x)\ ,
\end{equation}
where
\begin{eqnarray}
G_{\mu}(x)=\frac{e}{2}\ \left(  \frac{1}{3}+\tau^{3}\right) A_{\mu
}(x)+g_{W}\ \frac{1-\gamma_{5}}{2} \
\frac{\tau^+ W^+_{\mu}(x)+\tau^- W^-_{\mu}(x)}{\sqrt2}\ \ ,
\end{eqnarray}
with $g_{W}^{2}/(8M_{W}^{2})=G_{F}\,\cos\theta_{C}/\sqrt{2}$ and
$\tau^\pm = (\tau^1 \pm i \tau^2)/2$. In the present
case ---owing to the nonlocality of the involved fields--- one has to perform
additional replacements in the interaction terms, namely
\begin{align}
\psi(x-z/2)\  &  \rightarrow\ W_{G}\left(  x,x-z/2\right)  \ \psi
(x-z/2)\ ,\nonumber\\
\psi^{\dagger}(x+z/2)\  &  \rightarrow\ \psi^{\dagger}(x+z/2)\ W_{G}\left(
x+z/2,x\right)\ .  \label{gauge}%
\end{align}
Here $x$ and $z$ are the variables appearing in the definitions of the
nonlocal currents [see Eq.(\ref{cuOGE})], and the function $W_{G}(x,y)$ is
defined by
\begin{equation}
W_{G}(x,y)\ =\ \mathrm{P}\;\exp\left[  i\ \int_{x}^{y}dr_{\mu}\ G_{\mu
}(r)\right]  \ , \label{intpath}%
\end{equation}
where $r$ runs over an arbitrary path connecting $x$ with $y$.

Once the gauged effective action is built, the explicit expressions for the
vector and axial-vector form factors can be obtained by expanding to leading
order in the product $\delta\pi^{+} \, A_{\mu}\, W_{\nu}^{+}$. This is a
rather lengthy calculation that can be simplified by considering $q^2 \ll
m_{\pi^+}^{2}$, as needed to make predictions for the measured quantities
$F_{V}(0)$, $F_{A}(0)$ and $a$.

The vector form factor is obtained from the triangle diagram
represented in Fig.~1a. As stated, $F_{V}(q^{2})$ can be expanded at leading
order in $q^{2}$ as in Eq.~(\ref{fv}), just changing $q^2\to -q^2$ since we
are working in Euclidean space. Performing such an expansion we obtain
\begin{equation}
\frac{F_{V}(0)}{m_{\pi^+}}=\frac{\sqrt{2}\ \!g_{\pi q\bar q}\ \!N_{c}}{3}\int
\frac{d^{4}\ell}{(2\pi)^{4}}\ g_{\ell_{0}}\frac{(z_{\ell}+z_{\ell_{K}%
})(z_{\ell}+z_{\ell_{Q}})}{D_{\ell}\ D_{\ell_{K}}\ D_{\ell_{Q}}\ z_{\ell}%
}\left[  m_{\ell}-\frac{\ell^{2}}{2}\left(  \frac{m_{\ell_{Q}}-m_{\ell}%
}{Q\cdot\ell}-\frac{m_{\ell_{K}}-m_{\ell}}{K\cdot\ell}\right)  \right]  ,
\label{fv0}%
\end{equation}
where $K$ and $Q$ are the photon and $e^+\nu_e$ pair momenta, respectively,
taken at $q^2=0$. In Eq.(\ref{fv0})
we have used the definitions $D_{\ell}=\ell^{2}+m_{\ell}^{2}$,
$\ell_{Q}=\ell+Q$, $\ell_{K}=\ell-K$ and $\ell_{0}=(\ell_{Q}+\ell_{K})/2$.
For convenience we also define $\ell_{Q}^{\pm}=\ell\pm Q/2$,
$\ell_{K}^{\pm}=\ell\pm K/2$ and $\ell_{KQ}^{\pm}=\ell\pm (K+Q)/2$, which
will be used below.

The calculation of $a$ is somewhat more involved. In particular, the result
depends on the integration path appearing in the nonlocal contribution to
the quark$-$gauge boson vertices [see Eq.~(\ref{intpath})]. One obtains:
\begin{equation}
a=-\,\frac{m_{\pi^+}}{F_{V}(0)}\;\sqrt{2}\;\frac{N_{c}}{3}\,g_{\pi q\bar q}\int
\frac{d^{4}\ell}{(2\pi)^{4}}\ g_{\ell_{0}}\frac{(z_{\ell}+z_{\ell_{K}%
})\,(z_{\ell}+z_{\ell_{Q}})}{D_{\ell}\ D_{\ell_{K}}\ D_{\ell_{Q}}\ z_{\ell}%
}\ \mathcal{B}(\ell,K,Q)\ ,
\label{aaa}
\end{equation}
where
\begin{align}
\mathcal{B}(\ell,K,Q)  &  \ = \ \left[
m_{\ell}-\frac{\ell^{2}}{2}\left(  \frac{m_{\ell_{Q}%
}-m_{\ell}}{Q\cdot\ell}-\frac{m_{\ell_{K}}-m_{\ell}}{K\cdot\ell}\right)
\right] \nonumber\\
&  \hspace{-1cm} \times 2\;\bigg[K\cdot\ell_Q \left(  \frac{D_{\ell_{Q}}^{\prime}}%
{D_{\ell_{Q}}}-\frac{z_{\ell_{Q}}^{\prime}}{z_{\ell}+z_{\ell_{Q}}}\right)
+ K\cdot\ell_K \left(  \frac{D_{\ell_{K}}^{\prime}}{D_{\ell_{K}}}%
-\frac{z_{\ell_{K}}^{\prime}}{z_{\ell}+z_{\ell_{K}}}\right) - K\cdot\ell_0\;
\frac{g_{\ell_{0}}^{\prime}}{g_{\ell_{0}}}\bigg] \nonumber\\
&  \hspace{-1cm} +\ \ell^{2}\left(  \frac{K\cdot\ell_Q}{Q\cdot\ell}\ m_{\ell_{Q}%
}^{\prime}-m_{\ell_{K}}^{\prime}\right)  + \ \frac{1}{2}
\left(\frac{2\;K\cdot\ell}{K\cdot Q} - \frac{\ell^{2}}{K\cdot\ell}\right)
(m_{\ell_{K}}-m_{\ell})\nonumber\\
&  \hspace{-1cm}
+ \left(K\cdot\ell - \frac{\ell^{2}}{2}
\frac{K\cdot Q}{Q\cdot\ell} \right)%
\bigg[\frac{K\cdot\ell_Q}{Q\cdot\ell \;K\cdot Q}(m_{\ell_{Q}}-m_{\ell
})+\!\frac{2\,z_{\ell_{Q}}z_{\ell}}{z_{\ell_{Q}}+z_{\ell}}\left(
\!\alpha _{\ell,Q}+\frac{m_{\ell_{Q}}+m_{\ell}}{2}\beta_{\ell,Q}\right)
\!\bigg]\ .
\end{align}
Here primes stand for derivatives, e.g.~$g^\prime_\ell =
dg_\ell/d\ell^2$. The functions $\alpha_{\ell,Q}$ and
$\beta_{\ell,Q}$ are, in general, path dependent. Here. for simplicity,
we choose to use the ``straight line path" for which they read
\begin{equation}
\alpha_{\ell,Q}=\bar{\sigma}_{1}\int_{-1}^{1}d\lambda\ \lambda\ g_{\ell
_{Q}^{+}+\frac{\lambda}{2}\ Q}^{\prime}\ ,\qquad
\beta_{\ell,Q}=\bar{\sigma
}_{2}\int_{-1}^{1}d\lambda\ \lambda\ f_{\ell_{Q}^{+}+\frac{\lambda}{2}%
\ Q}^{\prime}\ .
\label{alfa}
\end{equation}

The axial-vector form factor receives not only a contribution from the
triangle diagram in Fig.~1a (as occurs in the local NJL model) but also from
other diagrams, which are represented in Figs.1b-1e. Thus, for $q^2=0$ one
has
\begin{align}
F_{A}(0) \ = \ \sum_{\alpha=a}^{e} F_{A}(0)|_{\alpha} \ .
\end{align}
The contribution from the triangle diagram is given by
\begin{align}
\frac{F_{A}(0)}{m_{\pi^+}}\Big|_{a}  &  = - \frac{\sqrt{2} g_{\pi qq} N_{c}%
}{(K.Q)^{2}} \int\frac{d^{4} \ell}{(2\pi)^{4}}\ \ g_{\ell_{0}} \ \frac
{z_{\ell_{Q}} \; z_{\ell_{K}}}{D_{\ell} D_{\ell_{Q}} D_{\ell_{K}}}
\ \ \mathcal{A}_{a} \ ,
\label{calat}
\end{align}
where
\begin{align}
\mathcal{A}_{a} \ = \ & 2 \left(  4 - \frac{K\cdot Q}{K\cdot\ell\ Q\cdot\ell
}\ \ell^{2} \right) \nonumber\\
& \times \, \Bigg\{ \! (\ell_{Q} \cdot\ell_{K} + m_{\ell_{Q}}\; m_{\ell_{K}})
\Bigg[ \! \left( \frac{m_{\ell^{+}_{Q}}}{z_{\ell^{+}_{Q}}} +
\frac{m_{\ell^{-}_{K}}}{z_{\ell^{-}_{K}}}%
- \frac{m_{\ell}}{z_{\ell}} \right)  D_{\ell} +
\, K\cdot Q\, \frac{ m_{\ell}\, ( z_{\ell_{Q}} + z_{\ell}) (z_{\ell
_{K}} + z_{\ell})} {4\;z_{\ell} z_{\ell_{Q}} z_{\ell_{K}} } \Bigg]\nonumber\\
&  - \frac{z_{\ell}\ m_{\ell^{-}_{K}}}{z_{\ell_Q} z_{\ell_K^-}} \;
(\ell\cdot\ell_{K} + m_{\ell}\;m_{\ell_{K}}) D_{\ell_{Q}} - \frac{z_{\ell
}\ m_{\ell^{+}_{Q}}}{z_{\ell_K} z_{\ell_Q^+}} \; (\ell\cdot\ell_{Q} +
m_{\ell}\;m_{\ell_{Q}}) D_{\ell_{K}} + \frac{ z_{\ell} \ m_{\ell}}{z_{\ell
_{K}}\;z_{\ell_{Q}}} D_{\ell_{Q}} D_{\ell_{K}} \Bigg\}\nonumber\\
&  + \ K\cdot Q\; \left(  2 - \frac{K\cdot Q}{K\cdot \ell\ Q\cdot\ell} \
\ell^{2} \right) \ \left(  1 + \frac{z_{\ell}}{z_{\ell_{Q}}}\right)  \left(
1 + \frac{z_{\ell
}}{z_{\ell_{K}}}\right) \nonumber\\
& \times\, \Bigg\{ \frac{m_{\ell^{+}_{Q}}}{z_{\ell^{+}_{Q}}}\
\frac{\ell \cdot\ell_{K} + m_{\ell}\;m_{\ell_{K}}} {1+
z_{\ell}/z_{\ell_{Q}}} \; + \; \frac{m_{\ell^{-}_{K}}}{z_{\ell^{-}_{K}}}\
\frac{\ell\cdot\ell_{Q} + m_{\ell }\;m_{\ell_{Q}}} {1+
z_{\ell}/z_{\ell_{K}}}\; -\; \frac{m_{\ell}}{z_{\ell}}\; (\ell_{0}^{2} +
m_{\ell_{Q}}\; m_{\ell_{K}}) \Bigg\}\ \! ,
\label{calaa}
\end{align}
while the remaining contributions are given by
\begin{align}
\sum_{\alpha=b}^e\frac{F_{A}(0)}{m_{\pi^+}}\Big|_\alpha
& = \frac{4
\sqrt{2}\, g_{\pi q\bar q}\, N_{c}}{(K\cdot Q)^{2}} \int\frac{d^{4}
\ell}{(2\pi)^{4}} \left[  \left( \frac{K\cdot Q}{K \cdot\ell\
Q\cdot\ell}\ \frac{\ell^{2}}{2} - 2 \right) \Big( \mathcal{A}_{b} + \mathcal{A}_{c} +
\mathcal{A}_{d} \Big) + \mathcal{A}_{e}\right]\ ,
\label{calar}
\end{align}
where
\begin{align}
\mathcal{A}_{b}  &  = \ ( g_{\ell}- g_{\ell^{-}_{K}} ) \left[  \frac
{z_{\ell^{+}_{Q}}\ m_{\ell^{+}_{Q}}}{D_{\ell^{+}_{Q}}}\ + \frac{z_{\ell
^{-}_{Q}}\ m_{\ell^{-}_{Q}}}{D_{\ell^{-}_{Q}}}\ - 2\;\frac{m_{\ell}}{z_{\ell}%
}\ \frac{z_{\ell^{+}_{Q}} z_{\ell^{-}_{Q}}}{D_{\ell^{+}_{Q}} D_{\ell^{-}_{Q}}}
\ \left(  \ell^{2} + \ m_{\ell^{-}_{Q}} \ m_{\ell^{+}_{Q}}\right)  \right]
\nonumber\\
\mathcal{A}_{c}  &  = g_{\ell^{-}_{K}} \ \Bigg[ \frac{z_{\ell^{+}_{Q}%
}\ m_{\ell^{+}_{Q}}}{D_{\ell^{+}_{Q}}} - \frac{z_{\ell^{-}_{Q}}\ m_{\ell
^{-}_{Q}}}{D_{\ell^{-}_{Q}}} \Bigg]\nonumber\\
\mathcal{A}_{d}  &  = \left(  g_{\ell} - g_{\ell^{-}_{K}}\right)
\Bigg[ \frac{z_{\ell}\ m_{\ell} }{D_{\ell}}- \frac{z_{\ell^{-}_{Q}}%
\ m_{\ell^{-}_{Q}} }{ D_{\ell^{-}_{Q}}}\Bigg]
\nonumber\\
\mathcal{A}_{e}  &  = \bar\sigma_1 \ g_{\ell}\ \frac{z_{\ell^{+}_{KQ}} \ z_{\ell^{-}_{KQ}}%
}{D_{\ell^{+}_{KQ}} D_{\ell^{-}_{KQ}}} \left(\ell^{+}_{KQ}\cdot
\ell^{-}_{KQ} + m_{\ell^{+}_{KQ}} \ m_{\ell^{-}_{KQ}}\right)
\left[  g_{\ell^{+}_{Q}}
+ g_{\ell^{-}_{K}} - g_{\ell_{0}} - g_{\ell} - \gamma(\ell,K,Q)\right]
\label{calar2}
\end{align}
with
\begin{align}
\gamma(\ell,K,Q) \ = \ & \frac{K\cdot Q}{2}\, \int_{0}^{1} \! d\lambda\
\frac{\left( \ell^{2} - 2\ \frac{K\cdot\ell\ Q\cdot\ell}{K\cdot Q}\right)
\left( g^{\prime}_{\ell+\frac{\lambda}{2}Q} -
g^{\prime}_{\ell^{-}_{K}+\frac{\lambda
}{2}Q}\right)  + \; g_{\ell+\frac{\lambda}{2}Q} \; - g_{\ell^{-}%
_{K}+\frac{\lambda}{2}Q}}{K\cdot\left(  \ell+ \frac{\lambda}{2} Q \right)  }
\ .
\label{gamma}
\end{align}
Note that, contrary to what happens with $F_{V}(0)$, the axial-vector form factor
depends on the path even at $q^{2}=0$. This is due to the contribution of the
diagram of Fig.~1e. The expression for $\gamma(\ell,K,Q)$ given above
corresponds to the ``straight line path'' choice.

Before discussing our predictions for the form factors associated with the
charged pion radiative weak decay, let us note that one can also consider
the related decay processes $\pi^{0}\rightarrow \gamma\gamma$ and
$\pi^{0}\rightarrow e^{+}e^{-}\gamma.$ The amplitude of these processes
contains the $\pi^{0}\gamma\gamma$ vertex form factor
$F^{\pi\gamma\gamma^{\ast}}\left(q^{2}\right)$, where $q^{2}$ is now the
invariant mass of the virtual photon. As in the case of $F_V(q^2)$ in the
decay $\pi^{+}\rightarrow e^{+}\nu\gamma$, one can perform an expansion
for low $q^{2}$:
\begin{eqnarray}
F^{\pi\gamma\gamma^{\ast}} \simeq
F^{\pi\gamma\gamma^{\ast}}(0)\left(1+a^{\prime}\ q^{2}/m_{\pi^{0}}^{2}
\right)\ , \qquad{\rm with}\qquad q^{2} \ll m_{\pi^{0}}^{2}\ .
\end{eqnarray}
Experimental measurements lead to $F^{\pi\gamma\gamma^{\ast}}(0)=0.284(8)$
$\mbox{GeV}^{-1}$ and $a^{\prime}=0.032(4)$~\cite{pdg}. Here it is
important to recall that, in the chiral limit, the anomaly leads to the
constraint $F^{\pi\gamma\gamma^{\ast}}(0) = 1/(4\pi^{2}f_{\pi}) \simeq
0.274 \ \mbox{GeV}^{-1}$. Moreover, using CVC and isospin arguments it is
easy to prove that $F_{V}(0) = m_{\pi}\,F^{\pi\gamma
\gamma^{\ast}}(0)/\sqrt{2}.$ We stress that our model satisfies this
relation as well as the anomaly constraint. In addition, it is easy to see
that it leads to an analytical expression for $a^{\prime}$ which coincides
with that given in Eq. (\ref{aaa}) for $a$.

We discuss now our numerical results. In Table I we list the predicted
values of $F_{V}(0)$ and the slope $a$ for the three parameterizations
considered in this work. We also include the available empirical data, as
well as the (local) NJL model predictions. In the case of $F_V(0)$, we
observe that the predictions of all three parameterizations of the
nonlocal model are in good agreement with the empirical value and with the
value obtained in the NJL model. This is hardly surprising, given the
chiral limit constraints mentioned above. Regarding the slope parameter we
find a too small value for all three parameterizations. There is a small
improvement by going from Set A to Set B (which implies the introduction
of the wave function renormalization of the quark propagator), and a
larger improvement is obtained by going from set B to C. One of the
peculiarities of the nonlocal model is that the results can be dependent
on the path used in Eq.~(\ref{intpath}). As usual, we have chosen a
straight line path for the calculations, obtaining the expressions given
in Eq.~(\ref{alfa}) for the path dependent quantities $\alpha_{\ell,Q}$
and $\beta_{\ell,Q}$. To gauge the importance of this path dependence we
have evaluated the contribution of the corresponding terms, obtaining that
it represents less than 3\% for parameterizations A and B and less than
0.5\% for Set C. As for the predictions for $a'$ we note that, even though
in our model its analytical expression coincides with that for $a$, the
corresponding numerical values are somewhat different since in the case of
$a^{\prime}$ we evaluate Eq.~(\ref{aaa}) at $P^{2}=m_{\pi^{0}}^{2}$,
whereas for $a$ it is evaluated at $P^{2}=m_{\pi^{+}}^{2}$. This dependence
on the mass of the pion reduces the value of $a^{\prime}$ in comparison
with that of $a$ by about 10\% for Sets A and B and 5\% for Set C.  We
observe in Table I that $a^{\prime}$ is better reproduced than $a$, but
the large errors in the experimental determination of the latter prevent
us to take definite conclusions. In fact, the discrepancies between our
predictions for $a$ and $a'$ and their experimental values are not
unexpected, since no vector interaction has been included in our model.
For example, using the results of Ref.~\cite{Prades:1993ys} we obtain that
the vector contribution to the $\pi^{0}\rightarrow\gamma \gamma^{\ast}$
process has the right order of magnitude and sign needed to account for
these discrepancies. Here it is interesting to point out that the
contribution from the vector channel could be different for $a$ and
$a^{\prime},$ even preserving the isospin symmetry. This would be achieved
if one has different interactions in the vector-isoscalar channel and in
the vector-isovector channel. On the other hand, it can be seen that the
NJL model is able to provide a very good prediction for $a^\prime$.
However, it is worth to notice that in the NJL framework this value is
essentially given by the relation $a^\prime\simeq
m_\pi^2/(12\,m_q^2)$~\cite{Noguera:2010fe}, and the value of $m_q$
(dressed quark mass) turns out to be quite dependent on model
parameterizations. The NJL values quoted here correspond to a
Pauli-Villars regularization.

In Table II we give our numerical results for $F_{A}(0)$, quoting the
contribution of each diagram. We have also quoted in Table II the result
obtained in the NJL model, where the triangle diagram is the only one that
survives. We observe that although several diagrams contribute in the case
of the nonlocal models, the triangle diagram turns out to be the dominant
one. The contribution coming from diagrams (b) to (e) is less than 2\% of
the total result, in all three parameterizations. In particular, diagram
(e), the only one with a path dependent term, contributes with less than 1\%
of the full result. It is interesting to mention that $F_{A}(0)$ has also
been evaluated in the spectral quark model, obtaining $F_{A}(0)\sim
F_{V}(0)$ as in the NJL model~\cite{Broniowski:2007fs}. In this sense, the
fact that nonlocal models of the type considered here lead to values of
$F_{A}(0)$ which are significantly different from those of $F_{V}(0)$
appears as a quite important result. A similar conclusion has been obtained
in a simplified calculation in the chiral limit~\cite{Kotko:2009ij}. Given
the triangle diagram dominance mentioned above, the origin of this
difference can be traced back to the different dressing of the
$\gamma_{\mu}$ and $\gamma_{\mu}\gamma_{5}$ terms in the coupling of the $W$
to the quarks \cite{Noguera:2008}. Comparing our prediction for $F_{A}(0)$
with the empirical value, we observe that the introduction of the
nonlocality gives half of the difference needed in the case of Set C and is
near to exhaust this difference for Set A, which gives the best result. In
any case, as in the case of the slope of the vector form factor, additional
effects that have been neglected in the present calculation might help in
improving the agreement with the empirical value.

\vspace*{.5cm}

We would like to acknowledge useful discussions with J. Portol\'es. This
work has been partially funded by the Spanish MCyT  (and EU FEDER) under
contract FPA2007-65748-C02-01, FPA2008-04810-E, FPA2010-21750-C02-01 and
AIC10-D-000588, by Consolider Ingenio 2010 CPAN (CSD2007-00042), by
Generalitat Valenciana: Prometeo/2009/129, by the European Integrated
Infrastructure Initiative HadronPhysics2 (Grant number 227431), by CONICET
(Argentina) under grants \# PIP 00682 and PIP 02495, and by ANPCyT
(Argentina) under grant \# PICT07 03-00818.

\vspace*{1cm}
\begin{table}[h]
\caption{Results for $F_{V}(0)$ and $a$. All results should be
multiplied by $10^{-2}$.
In column 5 we give the empirical values of $m_{\pi^+}\,F^{\pi\gamma \gamma^{\ast}}(0)/\sqrt{2}$
and $a'$. Note that in our model $a=a'$. In column 6 we give the prediction of the local
Nambu--Jona-Lasinio (NJL) model.
}
\begin{center}%
\begin{tabular}
[c]{ccccccc}\hline\hline
\hspace*{0.5cm}\ \hspace*{0.5cm} & \hspace*{0.5cm}Set A \hspace*{0.5cm} &
\hspace*{0.5cm}Set B \hspace*{0.5cm} & \hspace*{0.5cm}Set C \hspace*{0.5cm} &
\hspace*{0.5cm}Exp \cite{Bychkov:2008ws}\hspace*{0.5cm} & Exp ($\pi
^{0}\rightarrow\gamma\gamma^{\ast})$\cite{pdg} & \hspace*{0.5cm} NJL
\hspace*{0.5cm}\\\hline
$F_{V}(0)$ & 2.697 & 2.693 & 2.695 & 2.58 $\pm$ 0.17 & 2.80$\pm0.08$ & 2.441\\
$a$ & 1.651 & 1.726 & 2.011 & 10 $\pm$ 6 & 3.2$\pm0.4$ &
3.244\\\hline\hline
\end{tabular}
\end{center}
\label{tab1}%
\end{table}

\begin{table}[h]
\caption{Results for $F_{A}(0)$. All results should be multiplied by $10^{-2}$.
In column 5 we give the prediction of the local
Nambu--Jona-Lasinio (NJL) model.
}
\label{tab2}
\begin{center}%
\begin{tabular}
[c]{cccccc}\hline\hline
\hspace*{0.5cm}\ \hspace*{0.5cm} & \hspace*{0.5cm}Set A \hspace*{0.5cm} &
\hspace*{0.5cm}Set B \hspace*{0.5cm} & \hspace*{0.5cm}Set C \hspace*{0.5cm} &
\hspace*{0.5cm}Exp \cite{Bychkov:2008ws}\hspace*{0.5cm} & \hspace*{0.5cm} NJL
\hspace*{0.5cm}\\\hline
$F_{A}(0)|_{a}$ & 1.300 & 1.591 & 1.804 & & 2.409 \\
$F_{A}(0)|_{b}$ & 0.067 &  0.047 & 0.031 & & $-$ \\
$F_{A}(0)|_{c}$ & -0.0002 & -0.0002 & -0.0001 &  & $-$ \\
$F_{A}(0)|_{d}$ & -0.044 & -0.036 & -0.026 &  & $-$ \\
$F_{A}(0)|_{e}$ & -0.003 & 0.013 & 0.017 &  & $-$ \\\hline
$F_{A}(0)$ & 1.319 & 1.614 & 1.825 & 1.19 $\pm$ 0.01 & 2.409\\\hline\hline
\end{tabular}
\end{center}
\end{table}

\begin{figure}[tbh]
\includegraphics[width=0.8 \textwidth]{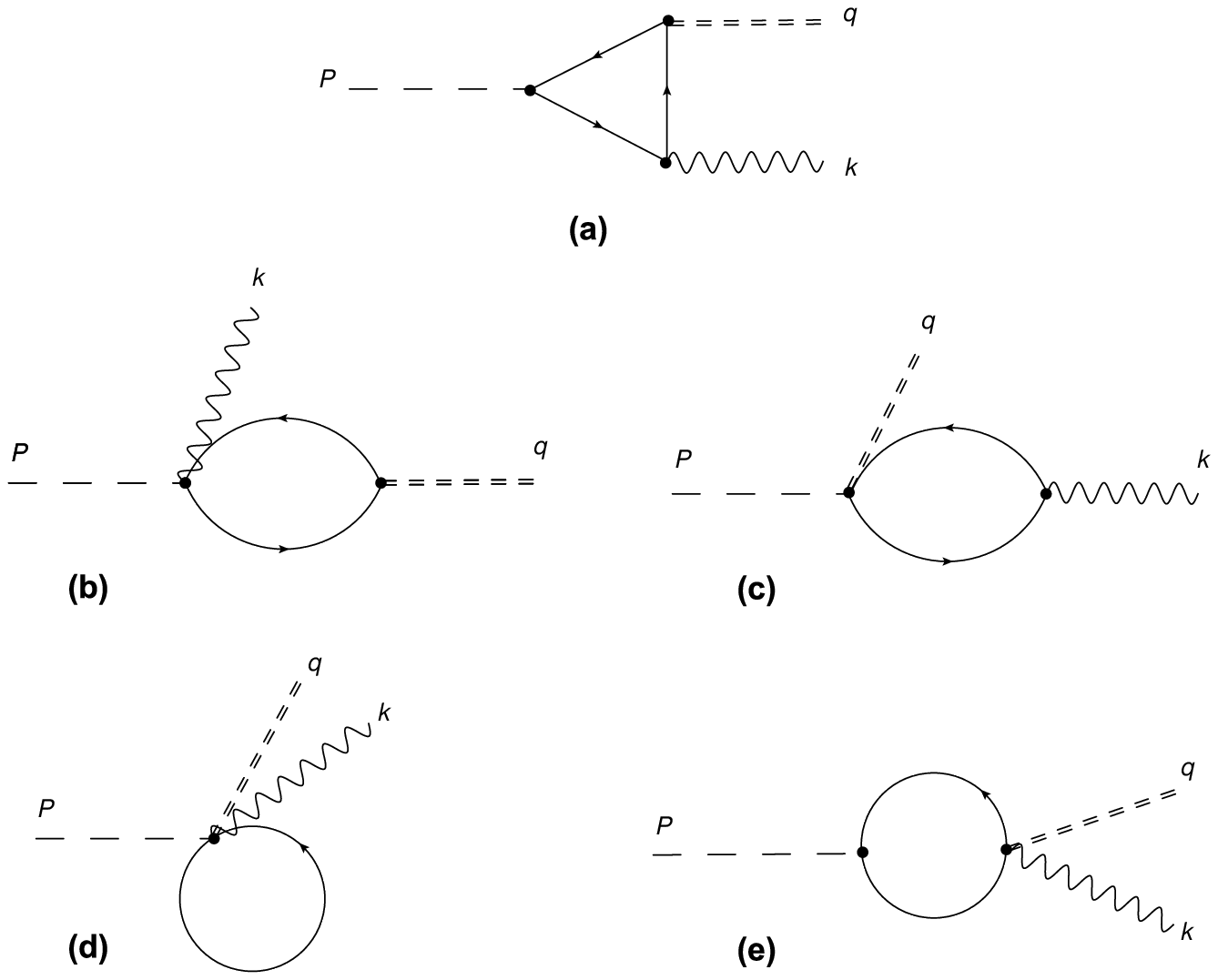}\caption{Diagrammatic
representation of the possible contributions to $\pi^{+}\rightarrow
e^{+}\nu_{e}\gamma$ decay. Double-dashed lines, wavy lines and
single-dashed lines represent the $e\,\nu_e$ pair, the outgoing photon and
the decaying pion, respectively. While for the vector form factor only the
contribution from the triangle diagram (a) is nonvanishing, in the case of
the axial-vector form factor all five diagrams contribute.} \label{diaI}
\end{figure}


\begin{thebibliography}{99}                                                                                               %

\bibitem {Bryman:1982et}
 M.~Moreno,
  Phys.\ Rev.\  D {\bf 16} (1977) 720;
D.~A.~Bryman, P.~Depommier and C.~Leroy,
Phys.\ Rept.\ \textbf{88} (1982) 151.


\bibitem {Bychkov:2008ws}M.~Bychkov \textit{et al.},
Phys.\ Rev.\ Lett.\ \textbf{103} (2009) 051802 [arXiv:0804.1815 [hep-ex]].

\bibitem{pdg}
K. Nakamura et al. (Particle Data Group),
J. Phys. G \textbf{37} (2010) 075021.

\bibitem {chiz}M. V. Chizhov,
  Mod.\ Phys.\ Lett.\  A {\bf 8} (1993) 2753
  [arXiv:hep-ph/0401217];
  Phys.\ Part.\ Nucl.\ Lett.\  {\bf 2} (2005) 193
  [Pisma Fiz.\ Elem.\ Chast.\ Atom.\ Yadra {\bf 2N4} (2005) 7]
  [arXiv:hep-ph/0402105].

\bibitem {pobla}
A. A. Poblaguev,
Phys. Rev. D \textbf{68} (2003) 054020 [arXiv:hep-ph/0307166].

\bibitem {Hol86}
B. R. Holstein,
Phys. Rev. D \textbf{33} (1986) 3316;
J.~Bijnens and P.~Talavera,
Nucl.\ Phys.\  B {\bf 489} (1997) 387
[arXiv:hep-ph/9610269];
  C.~Q.~Geng, I.~L.~Ho and T.~H.~Wu,
  Nucl.\ Phys.\  B {\bf 684} (2004) 281
  [arXiv:hep-ph/0306165].

\bibitem {Mateu:2007tr}
V.~Mateu and J.~Portoles,
Eur.\ Phys.\ J.\ C \textbf{52} (2007) 325 [arXiv:0706.1039 [hep-ph]].

\bibitem {Vogl:1991qt}U.~Vogl and W.~Weise,
Prog.\ Part.\ Nucl.\ Phys.\ \textbf{27} (1991) 195.

\bibitem {Klevansky:1992qe}S.~P.~Klevansky,
Rev.\ Mod.\ Phys.\ \textbf{64} (1992) 649.

\bibitem {Hatsuda:1994pi}T.~Hatsuda and T.~Kunihiro,
Phys.\ Rept.\ \textbf{247} (1994) 221.

\bibitem {Courtoy:2007vy}A.~Courtoy and S.~Noguera,
Phys.\ Rev.\ D \textbf{76} (2007) 094026 [arXiv:0707.3366 [hep-ph]].

\bibitem {Rip97}G.\ Ripka, \textit{Quarks bound by chiral fields} (Oxford
University Press, Oxford, 1997).

\bibitem {SS98}T.~Schafer and E.~V.~Shuryak,
Rev.\ Mod.\ Phys.\ \textbf{70} (1998) 323
[arXiv:hep-ph/9610451].

\bibitem {Roberts:1994dr}C.~D.~Roberts and A.~G.~Williams,
Prog.\ Part.\ Nucl.\ Phys.\ \textbf{33} (1994) 477
[arXiv:hep-ph/9403224];
C.~D.~Roberts and S.~M.~Schmidt,
Prog.\ Part.\ Nucl.\ Phys.\ \textbf{45} (2000) S1
[arXiv:nucl-th/0005064].

\bibitem {Parappilly:2005ei}M.~B.~Parappilly, P.~O.~Bowman, U.~M.~Heller,
D.~B.~Leinweber, A.~G.~Williams and J.~B.~Zhang,
Phys.\ Rev.\ D \textbf{73} (2006) 054504
[arXiv:hep-lat/0511007].

\bibitem {Bowman2003}P. O. Bowman, U. M. Heller, D. B. Leinweber and A. G.
Williams, Nucl. Phys. Proc. Suppl. \textbf{119} (2003) 323.
  [arXiv:hep-lat/0209129].
P. O. Bowman, U. M. Heller, and A. G. Williams, Phys. Rev. D \textbf{66}
(2002) 014505.
[arXiv:hep-lat/0203001].

\bibitem{Furui:2006ks}
  S.~Furui and H.~Nakajima,
  Phys.\ Rev.\  D {\bf 73} (2006) 074503.

\bibitem {Bowler:1994ir}R.~D.~Bowler and M.~C.~Birse,
Nucl.\ Phys.\ A \textbf{582} (1995) 655
[arXiv:hep-ph/9407336];
R.~S.~Plant and M.~C.~Birse,
Nucl.\ Phys.\ A \textbf{628} (1998) 607
[arXiv:hep-ph/9705372].

\bibitem {Scarpettini:2003fj}A.~Scarpettini, D.~Gomez Dumm and
N.~N.~Scoccola,
Phys.\ Rev.\ D \textbf{69} (2004) 114018
[arXiv:hep-ph/0311030].

\bibitem{GomezDumm:2006vz}
  D.~Gomez Dumm, A.~G.~Grunfeld and N.~N.~Scoccola,
  Phys.\ Rev.\  D {\bf 74} (2006) 054026
  [arXiv:hep-ph/0607023].

\bibitem {Golli:1998rf}B.~Golli, W.~Broniowski and G.~Ripka,
Phys.\ Lett.\ B \textbf{437} (1998) 24
[arXiv:hep-ph/9807261];
W.~Broniowski, B.~Golli and G.~Ripka,
Nucl.\ Phys.\ A \textbf{703} (2002) 667
[arXiv:hep-ph/0107139].

\bibitem {Rezaeian:2004nf}A.~H.~Rezaeian, N.~R.~Walet and M.~C.~Birse,
Phys.\ Rev.\ C \textbf{70} (2004) 065203
[arXiv:hep-ph/0408233];
A.~H.~Rezaeian and H.~J.~Pirner,
Nucl.\ Phys.\ A \textbf{769} (2006) 35
[arXiv:nucl-th/0510041].

\bibitem{Praszalowicz:2001wy}
  M.~Praszalowicz and A.~Rostworowski,
  Phys.\ Rev.\  D \textbf{64} (2001) 074003; 
  Phys.\ Rev.\  D \textbf{66} (2002) 054002 [arXiv:hep-ph/0111196].

\bibitem{Noguera}
S. Noguera, Int. J. Mod. Phys. E \textbf{16} (2007) 97
[arXiv:hep-ph/0806.0818]

\bibitem {Noguera:2005cc}S.~Noguera and V.~Vento,
Eur.\ Phys.\ J.\ A \textbf{28} (2006) 227 [arXiv:hep-ph/0505102].

\bibitem {Noguera:2008} S. Noguera and N. N. Scoccola,
Phys.\ Rev.\ D \textbf{78} (2008) 114002
[arXiv:hep-ph/0806.0818].

\bibitem{Noguera:2010fe}
   See e.g.\ S.~Noguera and V.~Vento,
  Eur.\ Phys.\ J.\  A {\bf 46} (2010) 197
  [arXiv:1001.3075 [hep-ph]].

\bibitem {Prades:1993ys}
J.\ Prades,
Z.\ Phys.\ C \textbf{63} (1994) 491 [Erratum-ibid. C \textbf{11} (1999) 571]
[arXiv:hep-ph/9302246].

\bibitem {Broniowski:2007fs} W. Broniowski and E. R. Arriola,
Phys. Lett. B \textbf{649} (2007) 49 [arXiv:hep-ph/0701243].

\bibitem{Kotko:2009ij}
  P.~Kotko and M.~Praszalowicz,
  Phys.\ Rev.\  D {\bf 80} (2009) 074002
  [arXiv:0907.4044 [hep-ph]].

\end{thebibliography}
\end{document}